\documentclass[11pt,twoside]{article}
\usepackage{asp2010}
\resetcounters

\begin{document}

\title{The HARPS polarimeter}
\author{Frans Snik$^1$*, Oleg Kochukhov$^2$, Nikolai Piskunov$^2$, Michiel Rodenhuis$^1$, Sandra Jeffers$^1$, Christoph Keller$^1$, Andrey Dolgopolov$^2$, Eric Stempels$^2$, Vitaly Makaganiuk$^2$, Jeff Valenti$^3$, and Christopher Johns-Krull$^4$
\affil{$^1$Sterrekundig Instituut Utrecht, Princetonplein 5, 3584 CC Utrecht, the Netherlands; *\texttt{\texttt{F.Snik@uu.nl}}}
\affil{$^2$Dept. of Physics and Astronomy, Uppsala University, Box 515, S-751 20 Uppsala, Sweden}
\affil{$^3$Space Telescope Science Institute, 3700 San Martin Drive, Baltimore, MD 21218, USA}
\affil{$^4$Rice University, Physics \& Astronomy Department, Houston, TX 77005, USA}}

\begin{abstract}
We recently commissioned the polarimetric upgrade of the HARPS spectrograph at ESO's 3.6-m telescope at La Silla, Chile. 
The HARPS polarimeter is capable of full Stokes spectropolarimetry with large sensitivity and accuracy, taking advantage of the large spectral resolution and stability of HARPS. 
In this paper we present the instrument design and its polarimetric performance. 
The first HARPSpol observations show that it can attain a polarimetric sensitivity of $\sim$10$^{-5}$ (after addition of many lines) and that no significant instrumental polarization effects are present. 
\end{abstract}

\section{Introduction}
The HARPS instrument \citep{HARPSpol-HARPS} attached to ESO's 3.6-m telescope at La Silla (Chile) has been very successful in detecting exoplanets through their radial velocity signatures.
This is mostly owing to the spectrograph's extreme stability.
Recently, HARPS' capabilities have been extended with a full-Stokes polarimeter, and it now furnishes high-resolution spectropolarimetric observations of stars and other bright objects like solar-system planets.
Spectropolarimetric observations with HARPSpol will be particularly interesting for measuring magnetic fields (including their spatial and temporal variations) on all kinds of stars.
By analyzing time-series of high-resolution spectropolarimetric data, one can map magnetic field properties on the surfaces of unresolved stars: Zeeman-Doppler imaging \citep[using only circular polarization;][]{HARPSpol-ZDI} or magnetic Doppler imaging \citep[using full-Stokes data;][]{HARPSpol-MDI}.\\

The new polarimeter takes full advantage of the two optical fibers that transport light from the Cassegrain focus to the stabilized spectrograph.
In regular HARPS observations, these fibers are fed by starlight and calibration light, respectively.
In the polarimetric mode, the fibers only contain starlight, after splitting according to perpendicular linear polarization directions.
In combination with a rotating quarter-wave plate and a half-wave plate as modulators, HARPSpol contains two dual-beam polarimeters (one for circular and one for linear polarization), that enable observations with a polarimetric sensitivity of $\sim$10$^{-5}$.
The polarimetric module now occupies the volume of the former Iodine cell, which was never used for HARPS observations.
A slider can insert the polarimeter from a completely retracted position, and allows for selection of the circular and the linear polarimeter.

\section{Requirements \& Design}

The following requirements were defined for the HARPS polarimeter \citep{HARPSpol-Sniketal2008}:

\begin{enumerate}
  \setlength{\itemsep}{1pt}
  \setlength{\parskip}{0pt}
  \setlength{\parsep}{0pt}

\item The full Stokes vector
$(I,Q,U,V)^T$ can be measured to enable complete diagnostics of magnetic field and scattering topologies. 
Because the 3.6-m telescope has an equatorial mount, the orientation of the $[Q,U]$ coordinate system is invariant on the sky.

\item The full wavelength range of HARPS (380--690 nm) is covered.

\item A polarimetric sensitivity of $\sim$$10^{-5}$ can be reached thanks to the beam-exchange technique \citep[see][]{HARPSpol-Semel1993,HARPSpol-Bagnuloetal2009}. In most cases this is only achieved after adding all lines with a technique like least-squares deconvolution \citep[LSD; ][]{HARPSpol-ZDI}, provided sufficient photon counts. Also the ``raw'' sensitivity is limited by photon noise.

\item The polarimetric accuracy needs to be as high as possible to furnish reliable interpretations of the data. 
The instrumental polarization of all the optics up until the polarimeter should therefore be minimized. 
Fortunately, the instrumental polarization at the Cassegrain focus is already minimal because the optical system is fully rotationally symmetric up to that focus, except for the atmospheric dispersion corrector (ADC), which has some small-angle refractions that cause up to 0.3\% of broad-band instrumental polarization and possibly some amount of stress birefringence that causes linear $\leftrightarrow$ circular polarization cross-talk.

\item The amplitudes of intensity and polarized fringes are minimized.

\item The polarimeter is designed such that the return beam to the guiding camera remains unblocked for all predefined positions of the slider (polarimeter out, circular polarimeter in, linear polarimeter in).

\end{enumerate}

\begin{figure}[t]

\includegraphics[width=0.97\textwidth]{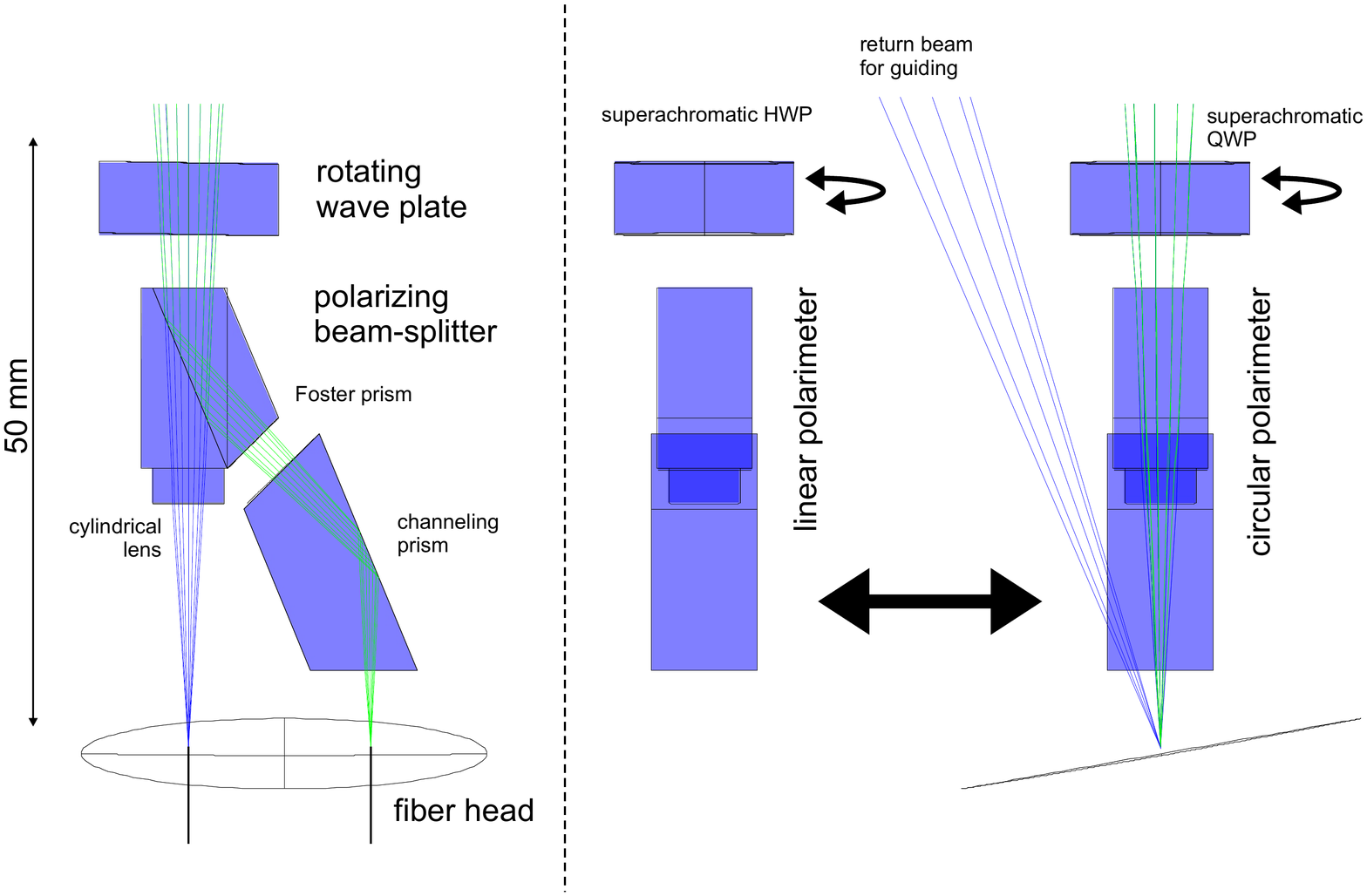}
\caption{Design of the HARPS polarimeter. The converging beam from the telescope is split by a custom polarizing beam-splitter and fed into the two HARPS fibers. The modulation is performed by a superachromatic QWP and HWP, respectively, for the two separate polarimetric units that can be positioned into the beam.}
\label{HARPSpol-design}
\end{figure}

The major design challenge for the HARPS polarimeter was to fit the optics for both polarimetric units into a small volume of $\sim$$5 \times 5 \times 10$ cm, see Fig.~\ref{HARPSpol-design}.
Because the spacing of the two fibers is 16 mm, no regular polarizing beam-splitter could be used.
Therefore we designed a custom beam-splitting assembly, based on a Foster prism (a modified Glan-Thompson prism) that yields an (achromatic) splitting angle of 135$^\circ$, with large extinction ratios ($10^4$--$10^6$) over the entire wavelength range.
The converging beam straight through the Foster prism suffers from crystal astigmatism, which is corrected by a cylindrical lens.
The second beam is injected into the second fiber by means of a total internal reflection inside a channeling prism that corrects for the focus difference between the two beams, and allows for precise alignment between the two beams.
The chromatic aberrations for both beams are within tolerances for feeding the fibers under seeing-limited conditions \citep{HARPSpol-Sniketal2008}.\\
\newpage
For the modulators, we opted for a superachromatic quarter-wave and half-wave plate (QWP, HWP), based on five layers of zero-order birefringent polymer \citep{HARPSpol-Samoylov2004}.
The deviations of the retardance value and the fast axis orientation from ideal values are measured to be within $\pm3^{\circ}$ and $\pm2^{\circ}$, respectively, for these plates over the entire wavelength range. 
These polymer wave plates have large acceptance angles and can therefore operate in the converging beam (contrary to Fresnel rhombs).
Furthermore, they barely generate any polarized fringes, which is a common problem with crystal wave plates.
To prevent any other kind of fringing, small wedge angles have been applied to all optical components.
Both wave plates are rotated simultaneously using a belt drive connected to an additional motor that is also located on the slider.
The amount of total defocus caused by the polarimetric optics is compensated by moving the telescope's secondary mirror backwards by $\sim$2 mm.\\

\begin{figure}[p]
\centering
\includegraphics[width=0.9\textwidth]{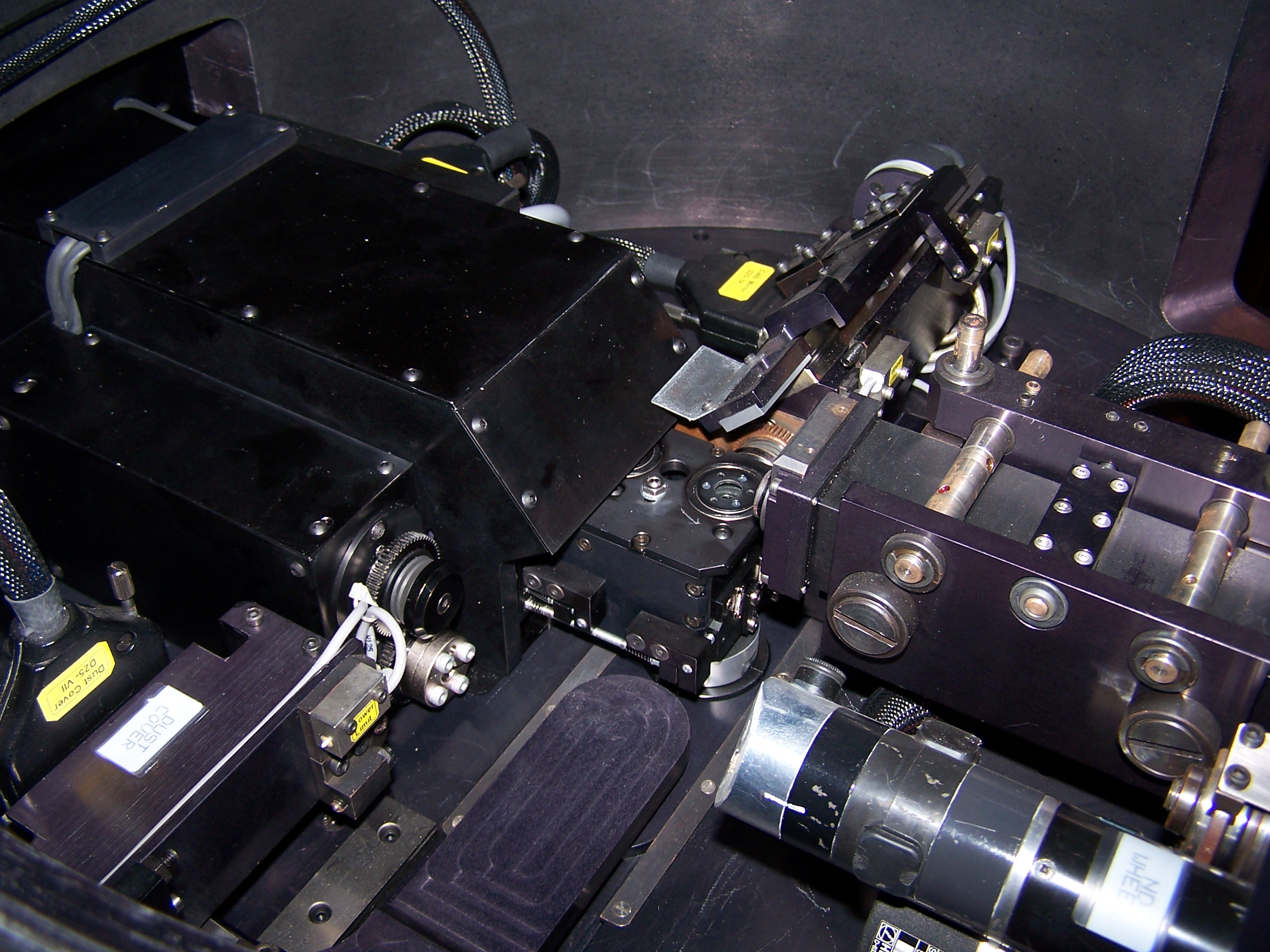}
\caption{Photo of the HARPS polarimeter after installation inside the Cassegrain adapter. The polarimetric module slides into the beam from under the ``garage'' on the left. In this case, the circular polarimeter with its rotating wave plate on top covers the fiber head.}
\label{HARPSpol-photo}
\end{figure}

\begin{figure}[p]
\centering
\includegraphics[width=0.9\textwidth]{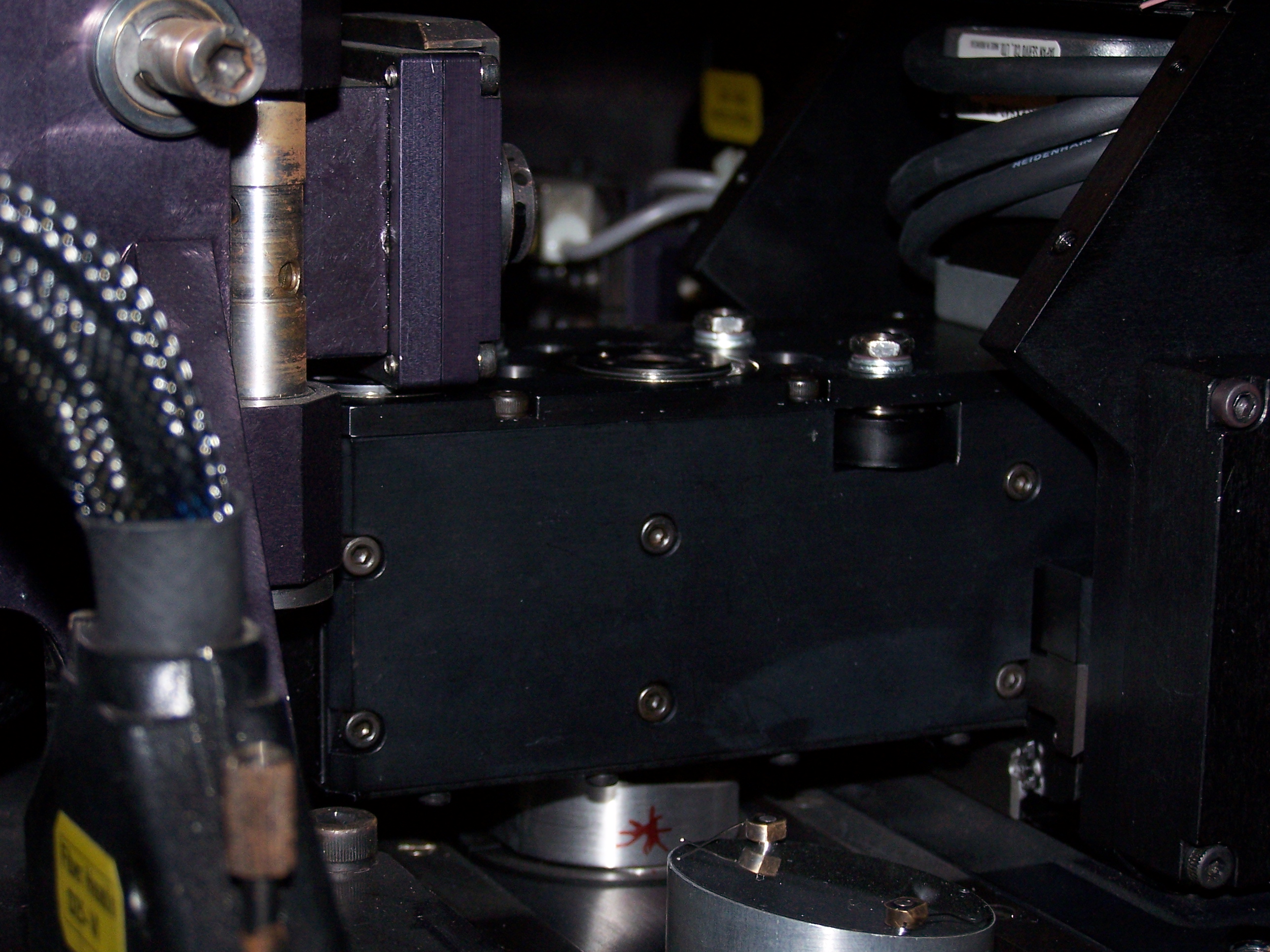}
\caption{Detailed view of the linear polarimeter sliding over the fiber head.}
\label{HARPSpol-photo2}
\end{figure}

The HARPS polarimeter was commissioned at the telescope in May 2009, and additional alignments were performed in December 2009.
See Figs.~\ref{HARPSpol-photo}, \ref{HARPSpol-photo2} for an impression of the polarimetric hardware inside the HARPS Cassegrain adapter. 

The two beams out of each beam-splitter have been aligned to the HARPS fibers such that the intensity ratios between the two beams as measured at the detector were matched to the ratio for the set-up without the polarimeter in to within 10\%.
The chromatic aberrations of the polarimetric optics were found to indeed be negligible during observations.

The wave plates were aligned with respect to the zero pulse of the encoder on the rotation mechanism.
Lab measurements using HeNe lasers with different wavelengths (543 and 633 nm) and two photodiodes allowed for the determination of the fast axis orientation with an accuracy of better than $0.3^\circ$.

At the Cassegrain adapter, HARPSpol was fully integrated with the HARPS instrument control electronics and software, which now contains two additional templates: ``circular polarimeter'' and ``linear polarimeter''.
The polarimetric unit itself fits in its designated space envelope to within 0.3 mm.\\

The linear polarimeter delivers measurements of $I \pm Q$ and $I \pm U$ in one beam, and $I \mp Q$ and $I \mp U$ in the other, by rotating the half-wave plate with incremental steps of 22.5$^\circ$.
The circular polarimeter yields measurements of $I \pm V$ (and $I \mp V$) by rotating the quarter-wave plate with 90$^\circ$ steps (from 45$^\circ$).
The redundancy in the polarization measurements in both beams at different positions of the modulator allows for elimination of differential effects (between the two beams due to limited transmission/gain calibration, and between the different modulation steps due to seeing or spectrograph drift), to first order, and thus increase the polarimetric sensitivity \citep{HARPSpol-Semel1993}.
Polarimetric data are demodulated using the ``double ratio'' method \citep[see][]{HARPSpol-Bagnuloetal2009}, and also null spectra can be determined to flag potential false signals.
All other standard data-reduction routines (e.g.~dark subtraction, flat-fielding, wavelength calibration) are performed using the standard HARPS pipeline.\\

The HARPS polarimeter is now available to the astronomical community as a standard observing mode of HARPS.
It offers similar capabilities as the ESPaDOnS instrument at the 3.6-m CFHT \citep{HARPSpol-ESPADONS}, although with larger spectral resolution ($R$ = \hbox{110 000}, compared to $R$ = \hbox{65 000}).
Furthermore, HARPSpol is unique in its coverage of the southern hemisphere.

\section{First Results}
Many different stars were observed during the HARPSpol commissioning and GTO runs.
The results of these observations will be presented in forthcoming publications.
Here, we present some representative data, to showcase HARPSpol's polarimetric sensitivity and accuracy.\\

\begin{figure}[t]
\centering
\includegraphics[width=0.9\textwidth]{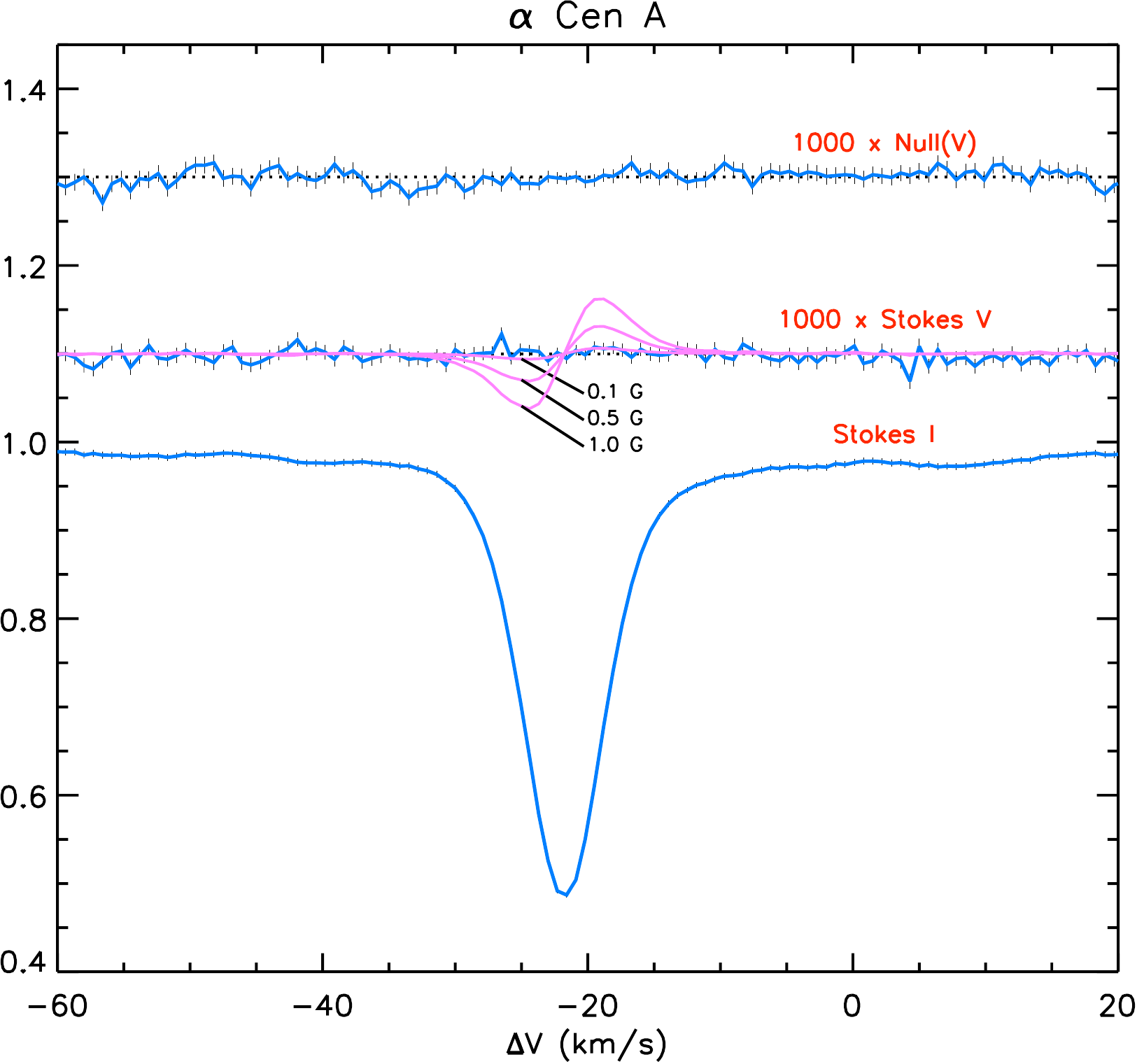}
\caption{HARPSpol observation of $\alpha$ Cen A. The LSD profile in Stokes $V$ has a noise level of 10$^{-5}$. No signal due to the Zeeman effect is detected down to a level that corresponds with a global magnetic field of 0.1 G.}
\label{HARPSpol-alphaCen}
\end{figure}

Commissioning data of $\alpha$ Cen A are presented in Fig.~\ref{HARPSpol-alphaCen}.
These data were obtained with the partially aligned circular polarimeter, and during cloudy conditions.
Nevertheless, sufficient photons were collected to reach a polarimetric sensitivity of $\sim$10$^{-5}$ for the LSD profile, which is a weighted average of all relevant spectral lines within HARPS' wavelength range.
No clear signals are detected in the LSD Stokes $V$ profile that correspond to a global stellar magnetic of more than 0.1 G.
In comparison, if a perfect solar twin had been observed, an average magnetic field due to active regions of $\sim$1 G would be expected \citep{HARPSpol-sunasstar}.
Furthermore, the Sun's polar region (latitudes larger than 70$^\circ$) is occupied by a unipolar magnetic with an average value \hbox{$>3$ G} \citep{HARPSpol-Tsunetapolar}, even during solar minimum.
Averaging over the entire visible disk reduces this value by an order of magnitude, but it is still larger than 0.1 G for most orientations of the polar axis with respect to the line of sight.\newpage
We therefore conclude that HARPSpol could in principle detect that a perfect solar twin is a magnetic star, and that $\alpha$ Cen A is significantly less magnetically active than the Sun, whereas they are both G2V stars of roughly the same age \citep[$\sim$4.9 Gy for \hbox{$\alpha$ Cen A} and \hbox{4.6 Gy} for the Sun;][]{HARPSpol-alphaCenage, HARPSpol-Sunage}.\\

\begin{figure}[p]
\centering
\includegraphics[width=\textwidth]{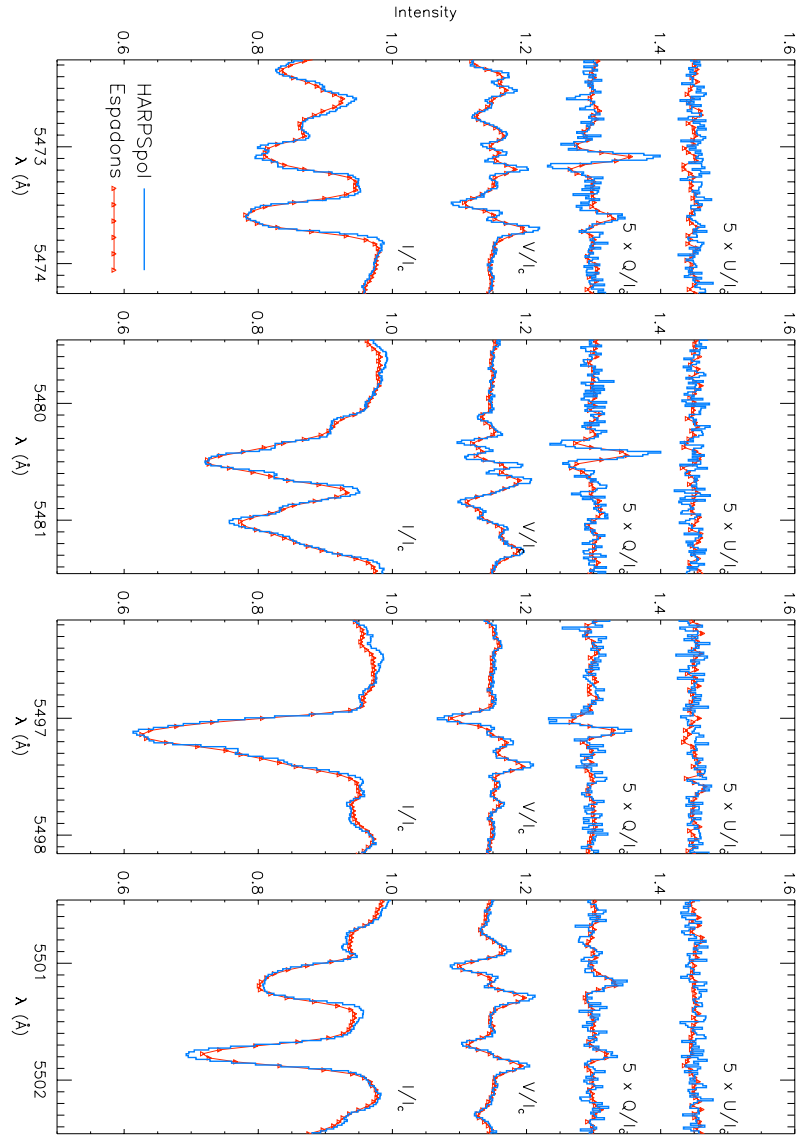}
\caption{Comparison of HARPSpol data with (calibrated) ESPaDOnS data of \hbox{$\gamma$ Equ}. The amplitudes in the HARPSpol data are larger owing to the larger spectral resolution of HARPS. The noise in $Q$ and $U$ is larger in the HARPSpol data due to imperfect alignment of the linear polarimeter at the time of the observations.}
\label{HARPSpol-gammaEqu}
\end{figure}

Unfortunately, polarization calibration optics could not be implemented in the design for HARPSpol.
To assess the polarimetric accuracy of HARPSpol, we observed the ``standard star'' $\gamma$ Equ, which is an Ap star with a very stable and very strong magnetic field \citep[$\sim$4 kG;][]{HARPSpol-gammaEquobs}.
Data for this target have also been obtained with ESPaDOnS, and have been rigorously analyzed to take out the effects of cross-talk.
The $\gamma$ Equ data from both HARPSpol and ESPaDOnS are presented in Fig.~\ref{HARPSpol-gammaEqu}.
It is clear that the uncalibrated HARPSpol data match well with the calibrated ESPaDOnS data, indicating the absence of significant cross-talk.
In general, the HARPSpol polarization spectra exhibit larger polarization signals, owing to its larger spectral resolution.
Broad-band instrumental polarization as induced by the ADC generally does not influence analyses based on line polarization.
Further efforts to quantify HARPSpol's instrumental polarization effects are ongoing.

\acknowledgements We thank the Geneva HARPS team and ESO staff for support of this project. We thank Gregg Wade for the ESPaDOnS data of $\gamma$ Equ.

\bibliography{Sniketal-HARPSpolSPW6}

\end{document}